\documentclass[runningheads,a4paper]{llncs}

\usepackage{amssymb}
\usepackage{graphicx}
\usepackage{algpseudocode}
\usepackage{enumerate}
\usepackage{amsmath}
\usepackage{xspace}
\usepackage{alltt}
\usepackage{authblk}
\usepackage{url}
\urldef{\mailsa}\path|{alfred.hofmann, ursula.barth, ingrid.haas, frank.holzwarth,|
\urldef{\mailsb}\path|anna.kramer, leonie.kunz, christine.reiss, nicole.sator,|
\urldef{\mailsc}\path|erika.siebert-cole, peter.strasser, lncs}@springer.com|

\newcommand{\keywords}[1]{\par\addvspace\baselineskip\noindent\keywordname\enspace\ignorespaces{#1}}
\newcommand{\remove}[1]{}
\newcommand{\laLDA}{{$\mathcal LA$}-LDA\xspace}
\usepackage{color}
\newcommand{\commentout}[1]{%
}

\urldef{\mailsa}\path{jeonhyuk@usc.edu,lerman@isi.edu}
\urldef{\mailsb}\path{getoor@cs.umd.edu}

\begin{document}

\mainmatter  

\title{
LA-LDA: A Limited Attention Topic Model for Social Recommendation
}

\author{\author{Jeon-Hyung Kang\inst{1} \and Kristina Lerman\inst{1} \and Lise Getoor\inst{2}}}
\institute{USC Information Sciences Institute, 4676 Admiralty Way, Marina del Rey, CA\\
\mailsa\\
\and
University of Maryland, Computer Science Department, College Park, MD\\
\mailsb\\
}

\maketitle
\begin{abstract}
Social media users have finite attention which limits the number of incoming messages from friends they can process.
Moreover, they pay more attention to opinions and recommendations of some friends more than others. In this paper, we propose \laLDA, a latent
topic model which incorporates limited, non-uniformly divided attention in the diffusion process by which opinions and information spread on the social network.
We show that our proposed model is able to
learn more accurate user models from users' social network and item
adoption behavior than models which do not take limited attention into
account. We analyze voting on news items on the social news aggregator Digg and show that our
proposed model is better able to predict held out votes than
alternative models. Our study demonstrates that psycho-socially motivated models
have better ability to describe and predict observed behavior than models
which only consider topics.


\end{abstract}

\section{Introduction}

Information overload has been drastically exacerbated by social media. On sites such as Twitter, YouTube and Facebook, more
videos and images are uploaded, blog posts written, and new messages posted than people are able to process. Social media sites attempt to mitigate this problem by
allowing users to subscribe to, or follow, updates from specific users
only. However, as the number of friends people follow grows, and the amount of information shared expands, the information overload problem returns.

Though social media contributes to the information overload problem; 
however it also 
creates opportunities for solutions. We can apply statistical techniques to social media data to learn user
preferences and interests from observations of their behavior. The
learned preferences could then be used to more accurately filter
and personalize streams of new information.
Consider social recommendation: when a user shares an item,
e.g., by posting a link to a news story on Digg or Twitter, he broadcasts it to all his followers.
Those followers may in turn
share the item with their own followers, and so on, creating a cascade through which information and
ideas diffuse through the social network. By analyzing these cascades,
who shares what items and when, we can learn what users are
interested in and use this knowledge to filter and rank incoming information.

The generic diffusion process described above ignores two important
elements: (\emph{i}) users have finite attention, which limits their
ability to process recommended items, and (\emph{ii}) users divide
their attention non-uniformly over their friends and
interests. Attention is the psychological mechanism that integrates
perceptual and cognitive factors to select the small fraction of input
to be processed in real
time~\cite{Kahneman73,Rensink:1997vj}. Attention has been shown to
be an important factor in explaining online
interactions~\cite{Weng:2012dd,Hodas12socialcom}.
Attentive acts, e.g., reading a tweet, browsing the web, or
responding to email, require mental effort, and since the brain's
capacity for mental effort is limited, so is attention. Attention has been
shown to impact the popularity of memes~\cite{Wu07,Weng:2012dd}, what
people retweet~\cite{Counts11,Hodas12socialcom} and the number of
meaningful conversations  they can have~\cite{Goncalves11}.
Attention is important, because most sites,
including Digg and Twitter, display items from friends as a
chronologically sorted list, with the newest items at the top of the
list. The more friends a user follows, the longer the list, in
average. A user scans the list, beginning at the top, and if he finds
an item interesting, he may share it with his followers. He will continue scanning the list until he gets bored or distracted, which is likely to happen before
he had a chance to inspect all new items.
While a user must divide his limited attention among his friends, he does not divide it uniformly. Some friends  are closer or more influential~\cite{Granovetter73,Gilbert09}; therefore, their recommendations may receive more attention, making them more likely to be adopted. Users may also preferentially pay more attention to each friend depending on topic.

In next section we describe a diffusion mechanism that takes into consideration the limited, non-uniformly divided attention of social media users. We use this mechanism to motivate \laLDA,  a probabilistic topic model we introduce.
Next, we analyze voting on news items on the
social news aggregator Digg and show that our model is better
able to predict held out votes than alternative models that do not
take limited attention into account.
Our study demonstrates that psycho-socially motivated models are better able to describe and predict observed user behavior in social media,
and may lead to better tools for solving the information overload problem.

\section{LA-LDA}

\subsubsection{Social Recommendation Setting}
\label{sec:diffusion}

We begin by describing the social recommendation scenario we are modeling.
We assume an idealized social media setting, with  $U$ users who recommend to each other and adopt items $A$.
Users have interests $X$, and items have topics $Z$, with users more likely to adopt items whose topics match their
interests. In addition, each user $u$ has $N_{frds(u)}$ friends and can see the items friends adopted.

The social recommendation model we propose is dynamic, and describes a number of user actions.
A user $u$ can share an item $i$ at time $t$.
An item could be  a link to an online resource that a user shares by tweeting it on Twitter or submitting for it on Digg.
We assume that when an item is shared by $u$, the recommendation is broadcast of all of $u$'s followers. A user $u$ can share a recommended item $i$ at time $t$, for example, by retweeting the link on Twitter or voting for it on Digg.

We also introduce the notion of a \emph{seed}, the user who introduced the item into the social network.  For any item $i$, there is a set of seed users whose adoptions diffuse through the social network along follower links, based on users'
interests.

Finally, what sets our model apart from previous models for social
recommendations is that we also model user's attention.  Users have limited attention and may not attend to all the items their friends recommend.  
After attending to an item, they may decide to adopt and share it. Once an item is shared, the limited attention diffusion process
continues to unfold.

In summary, in the context of social recommendation, limited attention
implies that users may process all items their
friends recommend. How they limit their attention
depends on both their interests and their social network.

\subsubsection{Probabilistic Model}
\label{ProbabilisticModel}
We now introduce a topic model \laLDA that captures the salient elements, including the limited attention of users, of social recommendation.
Our model consists of four key components which describe
user's interests ($\theta_{(u)}$),
item's topics ($\psi_{(i)}$),
user's attention to friends on different interests ($\tau_{(u)}$),
and user's limited attention ($\phi_{(u)}$).
We assume there are $N_u$ users, $N_i$ items, and each user $u$ follows
$N_{frds(u)}$ friends.
Moreover, each user has $N_x$ interests, and each item has $N_z$ topics. 

The \laLDA model is presented in graphical form in Figure 1(a).
There are four parts to the model
representation: user level ($\theta$, $\tau$, $\phi$), item level ($\psi$),
interest $\times$ topic level ($\pi$), and 
global hyperparameters ($\alpha, \beta, \rho$, and $\eta$).
Each adoption of an item $i$ by a user $u$ has an associated item topic $z$, and user interest $x$;  $Y$ denotes the friend(s) whose recommendations for $i$
were adopted by $u$.  Variables $A$ and $Y$ are observed, while $X$ and $Z$ are hidden.
User $u$'s interest profile $\theta_{(u)}$ is a distribution over $N_x$ interests.
Similarly, item $i$'s topic profile $\psi_{(i)}$ is a distribution over $N_z$ topics.
Each user pays attention to different friends depending on interests, so that for user $u$ and interest $x$, there is an interest-specific distribution $\tau_{(u,x)}$ over $frds(u)$.
The distribution of user $u$'s attention over both $N_x$ interests and $frds(u)$ is captured by $\phi_{(u)}$.
Finally, each interest $x$ and topic $z$ pair has an adoption probability $\pi_{(x,z)}$ for items. The generative process for item adoption through a social network is shown in Figure 1(b).

\begin{figure}[ht]
\begin{tabular}{cc}
\begin{minipage}[b]{0.4\linewidth}
\centering
\includegraphics[width=0.9\textwidth]{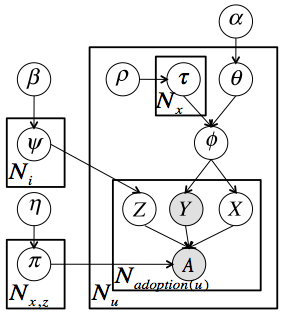}
\label{fig:modelDiagram}
\end{minipage}
&
\begin{minipage}[b]{0.55\linewidth}
\centering
\begin{alltt}
For each user u
  Generate\( \theta{(u)} \sim Dirichlet({\alpha}) \)
  For each interest x
    Generate\( \tau{(u,x)} \sim Dirichlet({\rho})\)
For each item i
  Generate\( \psi{(i)} \sim Dirichlet({\beta}) \)
For each interest x
  For each topic z
    Generate\( \pi{(x,z)} \sim Dirichlet({\eta}) \)
For each  user u
  For each adopted item i
    Choose interest x \(\sim Multinomial(\theta{(u)}) \)
    Choose friend to pay attention to y
                   \(\sim Multinomial(\tau{(u,x)}) \)
    Choose topic z\( \sim Multinomial(\psi{(i)}) \)
    Choose item i\( \sim Multinomial(\pi{(x,z)}) \)
\end{alltt}
\end{minipage}
\\ (a) & (b)
\end{tabular}
\caption{The \laLDA model (user interest profiles($\theta$), interest-specific attention profiles($\tau$), item topic profiles($\psi$), and adoption probabilities($\pi$)).}
\end{figure}

\subsubsection{Inference}
\label{sec:inference}
The inference procedure for our model follows the derivation of the
equations for collapsed Gibbs sampling, since we cannot compute
posterior distribution directly because of the summation in the
denominator. By constructing a Markov chain, we can sample
sequentially until the sampled parameters approach the target posterior
distributions. In particular, we sample all variables from their
distribution by conditioning on the currently assigned values of all
other variables. To apply this algorithm, we need the full conditional
distribution and it can be obtained by a probabilistic argument.

The Gibbs sampling formulas for the variables are:
\begin{equation}
\begin{aligned}
{P(Z_{(u,v)}=k|Z_{-(u,v)},X,Y,A_u) \propto \frac{n^{k}_{-(u,v)}+\beta}{n^{(\cdot)}_{-(u,v)}+\beta \times {N_z}} \frac{n^{x,k}_{-(u,v)}+\eta}{n^{x,k}_{-(\cdot,\cdot)}+\eta \times {N_i}} }\\
{P(X_{(u,v)}=j |X_{-{(u,v)}}, Y, Z, A_u) \propto       \ \  \  \  \ \ \  \  \ \ \  \  \ \ \  \ \  \ \ \  \  \ \ \  \  \ \ \  \  \ \ \  \  \ \ \  \  \ \ \  \  \ \ \ }
\\ \frac{n^{j}_{-(u,\cdot)}+\alpha}{n^{(\cdot)}_{-(u,\cdot)}+\alpha\times {N_x}} \frac{n^{y}_{-(u,j)}+\rho}{n^{(\cdot)}_{-(u,j)}+\rho \times N_{(frds(u))}}\frac{n^{j,z}_{-(u,v)}+\eta}{n^{j,z}_{-(\cdot,\cdot)}+\eta \times {N_i}}
 \end{aligned}
\end{equation}

\noindent where $n^{k}_{-(u,v)}$ is the number of times topic $k$ is assigned on
item $(u,v)$ excluding the current assignment of
$Z_{(u,v)}$, $n^{x,k}_{-(u,v)}$ is the number of adoptions of item
$(u,v)$ under item topic assignment $k$ and user interest assignment
of $x$, excluding the current item topic assignment of
$Z_{(u,v)}$,
$A_u$ is the set of items adopted by user u,
and $v$ ranges over the items in $A_u$. $(u,v)$ denotes
the index of the $v$th item adopted by user $u$.
The first ratio expresses the probability of topic $k$
for item $(u,v)$, and the second ratio expresses the probability of
item $(u,v)$'s adoption under the item topic assignment $k$ and user
interest assignment $x$. In the second equation, $n^{j}_{-(u,\cdot)}$
is the number of times user $u$ pays attention to interest $j$
excluding the current assignment of $X_{(u,v)}$ and $n^{y}_{-(u,j)}$
is the number of times user $u$ pays attention to friend $y$ on
interest $x$ excluding the current assignment of $X_{(u,v)}$. The
first ratio expresses the probability of user $u$ paying attention to
interest $j$ and the second ratio expresses the probability that user
$u$ pays attention to friend $y$ on interest $j$.
Our model
allows the algorithm learn each user's interests by taking into account
the limited attention on friends for certain interests from local perspective, while adopting is given by
user's interest and item's  topic assignment from global
perspective. To make the model simple we use symmetric Dirichlet priors.
We estimate $\theta$, $\psi$, $\pi$, and $\phi$ with sampled values
in the standard manner.

\section{Evaluation on  Synthetic Data}

Our first set of experiments illustrate the properties of the
\laLDA model used in conjunction with synthetic data. 
We used social network links among top 5,000 most active users in 2009 dataset, who are followed by in average 81.8 other users (max 984 and median 11).
We begin generating synthetic data by creating $N_i$ items and $N_u$ users according to the generative model.

We model the propagation of items through the social network over a
period of $N_{day}$ days.  We first choose a set of seeders ($S \%$) from  $N_u$ users.  Seeders will be able to introduce new items into
the network.  We introduce a special source node, which contains all
of the items.  Seeders will have the source node as one of
their friends.
Every user $u$ is assigned a fixed attention budget
$V_{u}$, which determines the total number of items from friends that $u$ can
attend to in a day.
For simplicity, we represent $V_{u}$ as a function of
a global attention limit parameter $v_g$ and the
number of friends user has. This is motivated by the observation that,
at least on Digg, user activity is correlated with
the number of friends they follow (the correlation coefficient is
0.1626--0.1701). Intuitively, the number of items
a user adopts is some fraction of the number of stories to which a user attends; here, to simplify matters,
we assume that user's attention budget is simply proportional to the number of friends
she follows.
\begin{algorithmic}
\Function{Generate Synthetic Data}{} 
\For{$day = 1 \to N_{day}$}
   \For{$u = 1 \to N_u$}
	\For{$attention = 1 \to V_{u}$}
 	    \State choose interest $x \sim Mult(\theta_{(u)})$
 	    \State choose friend $y \sim Mult(\tau_{(u,x)})$
		\State choose a item $i$ from $y$ 
	    \State  choose topic $z  \sim Mult(\psi_{(i)}$)
	    \State Adopt and share item with probability $\pi_{(x,z)}$
	\EndFor
	\EndFor
\EndFor
\EndFunction
\end{algorithmic}

Synthetic cascades are generated as follows.
Each day, every user within her allotted attention budget, will check to see whether her friends have
any items that match her interests.   Initially, when the cascade starts, the source node is the only
friend, which has items, so only seed nodes will be able to adopt and share items.  However, as time progresses,
and items begin flowing through the network. Eventually users will exhaust their attention budget, without
being able to attend to all the items that their friends shared with them.  When user
chooses to attend to an item $i$ that has been shared by a friend $y$, they choose without replacement, so that an
item will only be attended to once from a particular friend $y$.   However, we do allow a user to attend
the same item from different friends.  Once an item has been chosen, the user will adopt (and share) the
item with probability $\pi_{x,z}$.  

By varying parameters ($S$ and $v_g$) and hyperparameters ($\alpha$, $\beta$, $\eta$, and $\rho$) we can create different synthetic datasets and  we investigate how well we are able to recover
the user interests from the generated data using \laLDA (or LDA) model. 
We evaluate the performance of models by measuring the similarity of the learned and the actual distributions by the average
deviation between the Jensen-Shannon divergence of their
vectors. The average deviation is small when two vectors are similar without considering the indexing of the interests.  

\begin{figure}[tbh]
\begin{center}
\begin{tabular}{cccc}
\includegraphics[width=0.24\linewidth]{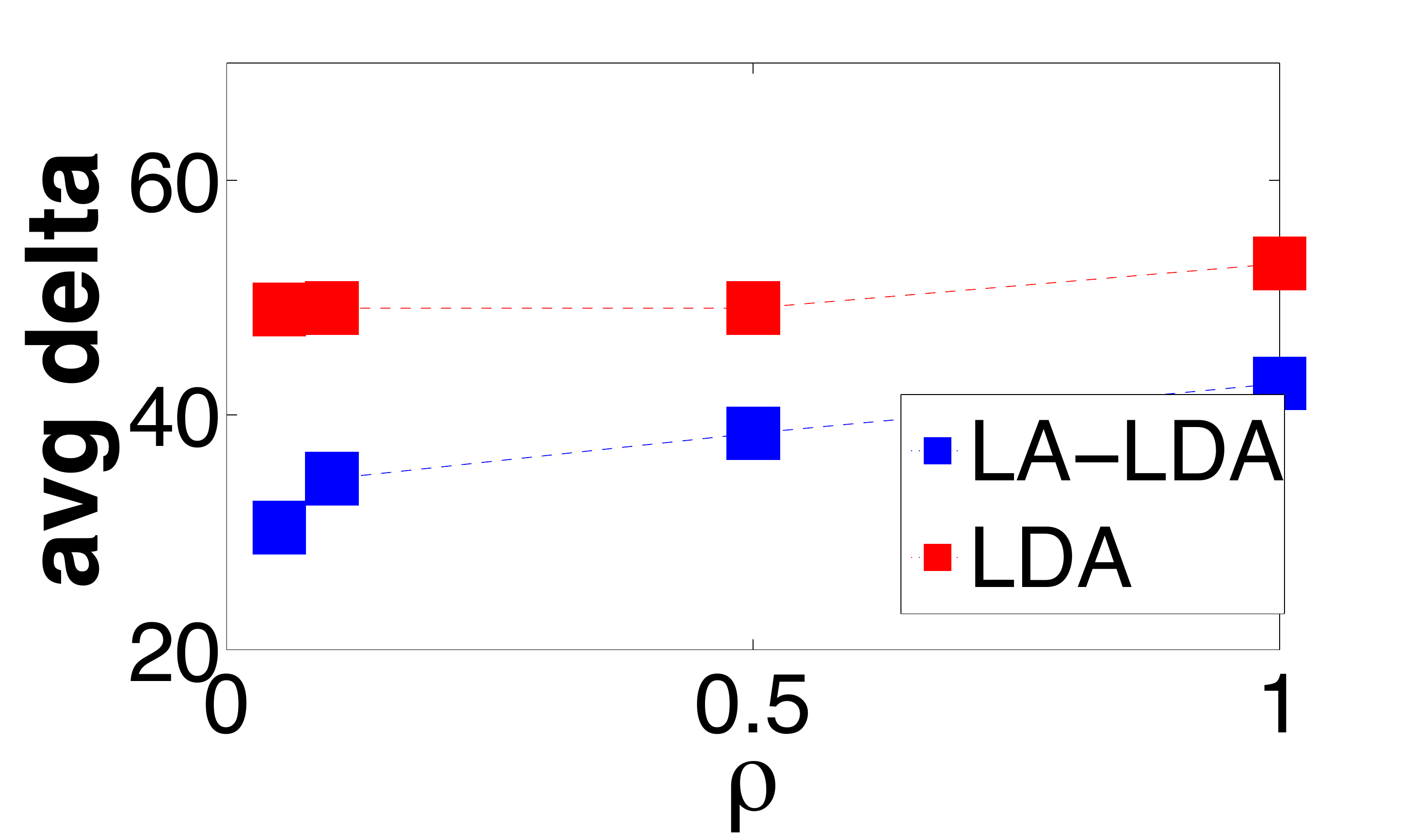}&
\includegraphics[width=0.24\linewidth]{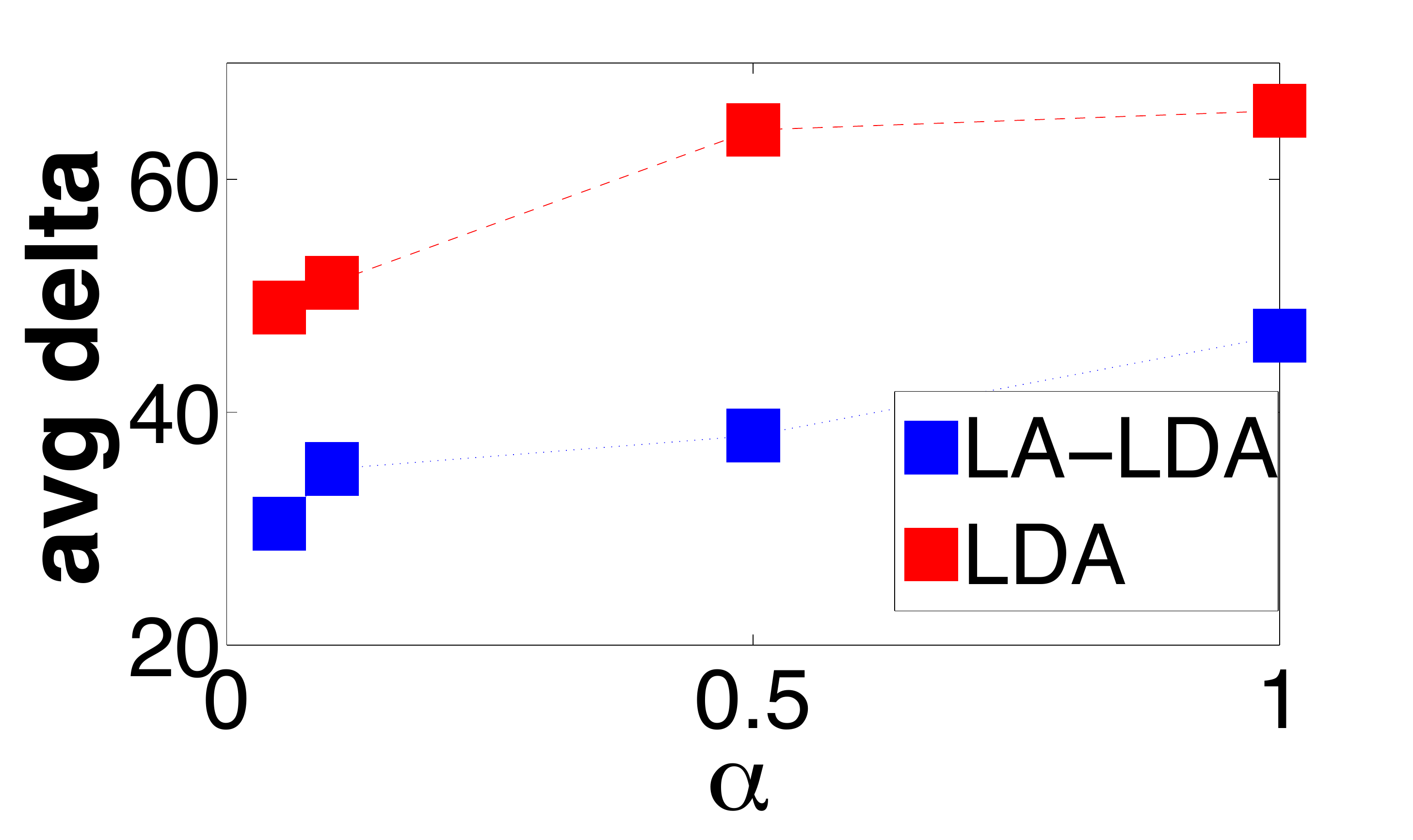}&
\includegraphics[width=0.24\linewidth]{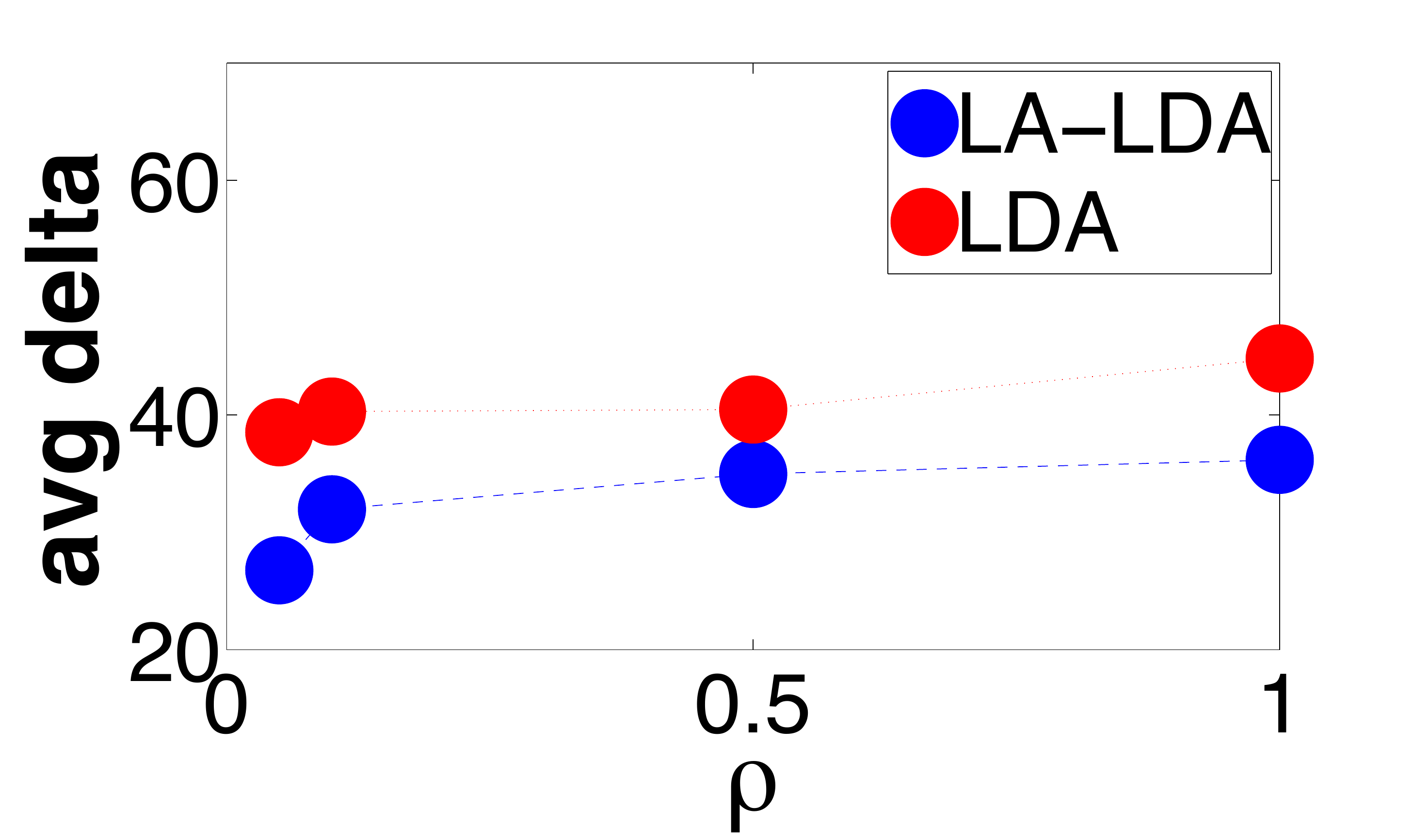}&
\includegraphics[width=0.24\linewidth]{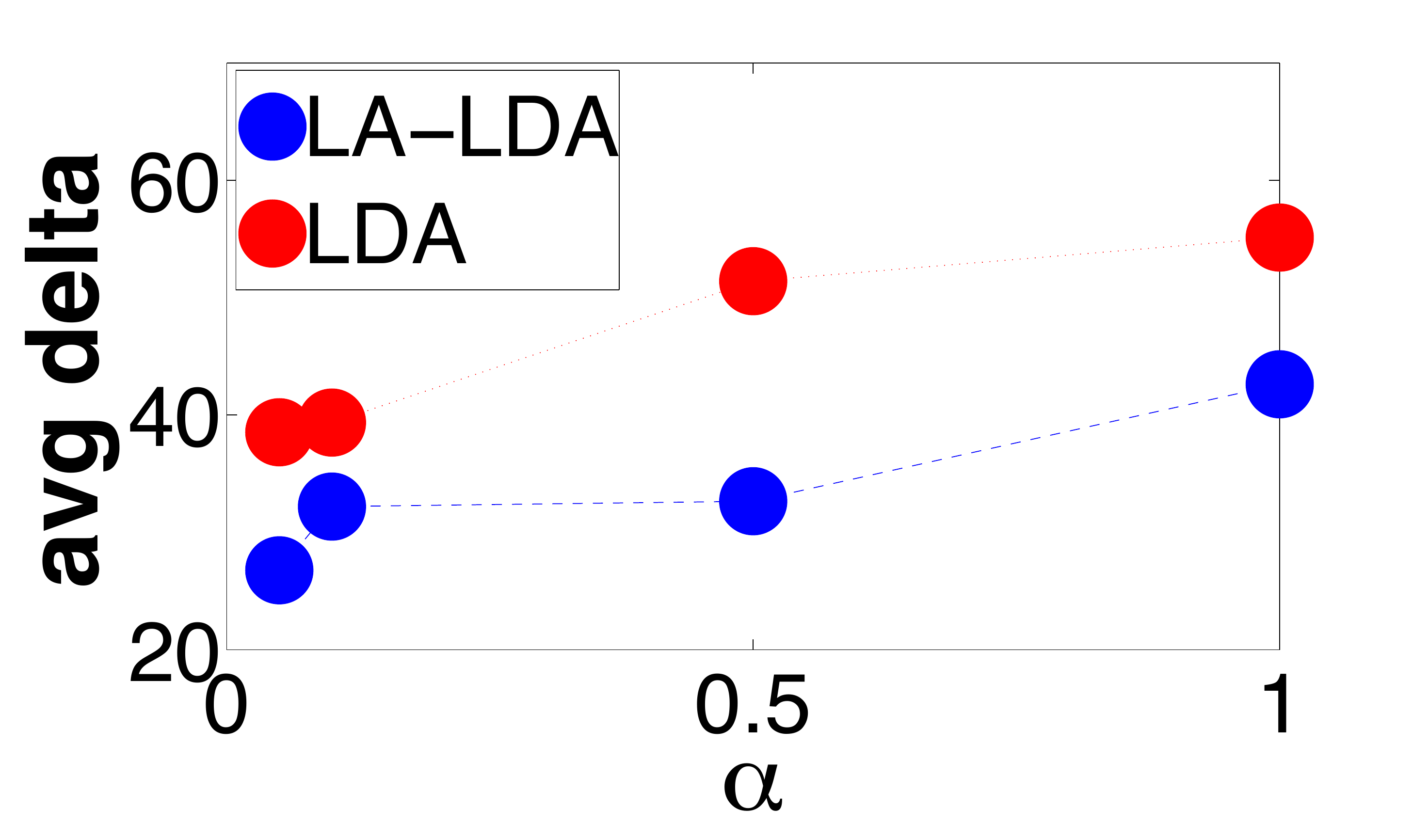}\\(a)&(b)&(c)&(d)
\end{tabular}
\end{center}
\caption{The average deviation of user interest ($\theta$) and item topic ($\psi$) with different limited attention values ($\rho$ and $\alpha$) on synthetic. The top two figures show average deviation between learned and actual $\theta$ when (a) $\alpha$=0.05 and $\rho$=0.05, 0.1, 0.5, and 1.0 and (b) $\rho$=0.05 and $\alpha$=0.05, 0.1, 0.5, and 1.0. The bottom two figures show average deviation between learned and actual $\psi$ when (c) $\alpha$ = 0.05 and  (d) $\rho$ = 0.05.  }
\label{fig:avgdeltaSyn}
\end{figure}
For comparison, we learned two different LDA models,
one for user interests and one for item topics. 
We learn the LDA for interest distributions of users $\theta$ by
viewing a user as a document and items as terms in a document, and we learn the LDA for topic
distributions of items $\psi$ by setting item as a document and users
as terms in a document. We also ran \laLDA to learn both $\theta$ and
$\psi$ in accordance with that model.  
For generating the synthetic data,
we set
$v_g$=2, $\beta$=0.1, $\eta$=0.1 and $S$=30$\%$) and varied
$\alpha$ (0.05, 0.1, 0.5, and 1.0) and $\rho$ (0.05, 0.1, 0.5, and
1.0). We applied the same hyperparameters used to
generate the synthetic data in the models.

The average
deviation between learned and actual interests and topics
of items 
in the synthetic datasets are shown in
Fig.~\ref{fig:avgdeltaSyn}.
With large values of $\alpha$, users
allocate their attention uniformly over interests, so users are more
likely to adopt items on a variety of interests. Because of this adoption
tendency, it is hard to distinguish
their interests. For small values of $\alpha$,
users pay attention to a limited number of interests and more
can be learned from their adoption behavior.
That is why both LDA and \laLDA perform
better for small $\alpha$ values.
Similarly, large values of $\rho$ cause users to
pay attention to their friends uniformly, while small values focuses 
users' attention to a smaller subset of their friends. With large $\rho$
values, average deviations of both models are high,
whereas for lower values both models perform better.
In all four cases, \laLDA
is superior to LDA in learning interests distribution of users and
topics distribution of items for all $\alpha$ and $\rho$ values.

\section{Evaluation on Digg}
\label{sec:eval}
We evaluate \laLDA on real-world data from the
social news aggregator Digg, which allows users to submit
links to news stories and other users to vote for (or ``digg'')
stories they find interesting.
Digg also allows users to follow the activity of other
users to see the stories they submitted or dugg recently. When a user votes for a story,
this recommendation is broadcast to all his followers.
At the time data was collected, users were submitting many thousands of stories, from which Digg selected a handful to promote to its popular front page.

We evaluated two datasets The 2009
dataset~\cite{Lerman10icwsm} contains information about the voting history of
70K active users (with 1.7M social links) on 3.5K stories promoted to Digg front page in June,
and contains 2.1M votes.
 At the time, Digg
assigned stories to one of eight topics (Entertainment,
Lifestyle, Science, Technology, World \& Business, Sports, Offbeat,
and Gaming).
The 2010 dataset~\cite{sharara:icwsm11} contains information about voting histories of
12K users (with 1.3M social links) over a 6 months period (Jul -- Dec). It includes
48K stories with 1.9M votes.  At the time data was collected, Digg assigned stories to
10 topics, replacing the ``World \& Business'' topic with ``World News,''
``Business,'' and ``Politics''.

Before a story is promoted to the front page, it is visible on the
upcoming stories queue and to the submitter's followers.
With each new vote, the story becomes visible to that voter's
followers. We examine only the votes that the story accrued before promotion to
the front page, during which time it propagated mainly via friends' recommendations.
In the 2009 dataset, 28K users
voted for 3K stories and in the 2010 dataset, 4K users voted for 36K stories
before promotion. We focused the data further by selecting those users who voted at
least 10 times, resulting in 2,390 users (who voted
for 3,553 stories) in the 2009 dataset and 2,330 users (who voted on
22,483 stories) in the 2010 dataset.

\laLDA has six parameters: the number of interests ($N_x$) and topics ($N_z$) and hyperparameters $\alpha$, $\beta$, $\eta$, and $\rho$. The choice of hyperparameters can have implications inference results. While our algorithm can be extended to learn hyperparameters, here we fix them (0.1)
and focus on the consequences of varying the number of topics and interests (from 5 to 800).
We estimate the performance of model by computing the likelihood
of the training set given the model for different combinations of parameters.
We took samples at a lag of 100 iterations after discarding the first
1000 iterations and both algorithms stabilize within 2000 iterations.
The best performance is obtained for
$N_x=10$ interests and $N_z=200$ topics in the 2009 dataset and
$N_x=30$ interests and $N_z=200$ topics in the 2010 dataset for both ITM
and \laLDA.
LDA results in best performance for 200 interests in the 2009 and 500 interests in the 2010 dataset.

\subsubsection{Evaluation of Learned User Interests}

The topics assigned to stories by Digg provide useful evidence for
evaluating topic models. We represent user $u$'s preferences by
constructing an empirical interest vector that gives the fraction
of votes made by $u$ on each topic. The empirical interest vector
serves as gold standard for evaluating user interests learned by
different topic models. We measure the similarity of the distributions using average
Jensen-Shannon divergence. In both datasets, \laLDA
(2009 dataset: 15.11 \& 2010 dataset: 28.71) outperforms ITM~\cite{Plangprasopchok11tkdd}  (36.38 \& 36.01) and LDA~\cite{blei2003latent} (37.72 \& 55.43)  models by learning user interests that are closer to the gold standard.

\subsubsection{Evaluation on Vote Prediction}

We evaluate our
proposed topic models by measuring how well they allow us to predict individual votes. There are
257K pre-promotion votes in the 2009 dataset and
1.5M votes in the 2010 dataset, with 72.34 and 68.20 average
votes per story, respectively.  For our evaluation, we randomly split
the data into training and test sets, and performed five-fold cross validation.
To generate the test set, we
use the held-out votes (positive examples) and
augment it with stories that friends
of users shared but that were not adopted by user. 
Depending on a user's and their friends' activities, there are different numbers of positive ($N_{pos}^{u}$)  in the test set. 
The
average percentage of $N_{pos}^{u}$ in the test set is 0.73\% (max 18\%, min 0.02\%, and median 0.13\%), suggesting that friends share many stories that users do not end up not voting for.
This makes the prediction task extremely challenging, with less than one in a hundred chance of
successfully predicting votes if stories are picked randomly.

We train the models on the data in the training set. Then, for each story $i$ in the test set, we compute the probability user $u$ votes for it, given training data $\mathcal{D}$. 
For LDA, the probability of the vote on $i$ is the probability of adopting $a_i$:
\begin{equation}
\label{eq:predict-lda}
\begin{aligned}
P(a_i|\mathcal{D}) =
      \int_{\theta} \sum_{x} P(a_i | x) P(x | \theta) P(\theta | \mathcal{D})\,\mathrm{d}\theta
\end{aligned}
\end{equation}
For  ITM, the probability that user $u$ votes for story $i$ is obtained by integrating over the posterior Dirichlet distributions of $\theta$ and $\psi$:
\begin{equation}
\label{eq:predict-itm}
\begin{aligned}
P(a_i|\mathcal{D}) =
     \int_{\psi} \int_{\theta} \sum_{x,z} P(a_i | {z,x})P(z | \psi) P(x | \theta) P(\psi | \mathcal{D})P(\theta | \mathcal{D}) \, \mathrm{ d\theta d\psi}
\end{aligned}
\end{equation}
\noindent Finally, in the \laLDA model, the probability user $u$ votes for  story $i$ is:
\begin{equation}
\label{eq:predict-lalda}
\begin{aligned}
     P(a_i|\mathcal{D}) =
     \int_{\psi} \int_{\phi} \sum_{x,y,z} P(a_i | {x,z})P(z | \psi) P(x,y | \phi)P(\psi | \mathcal{D})P(\phi | \mathcal{D}) \, \mathrm{ d\phi d\psi}
\end{aligned}
\end{equation}
\noindent where the probability of a user's vote is decided by the distribution of the user's limited attention over friends and interests $\phi$ and story's topic profile $\psi$.
We evaluate performance of the models on the prediction task using
average precision.
Average precision at $N_{pos}^{u}$ for each user is $\sum_{k=1,n} Prec(k) / (N_{pos}^{user}) $,
\noindent where $Prec(k)$ is the precision at cut-off $k$ in the list of votes ordered by their likelihood.

We divide users into categories based on their activity in the training set. The first category
includes \emph{all users} and the remaining categories include users
who voted for at least 7.5\%, 15\%, and 25\% of the stories in the training set.
While \laLDA outperforms baseline
methods in all cases, its comparative advantage improves with user
activity. When there is little information about user interests,
the precision of all methods is ranges from 1\%--3\%. As the amount of
information about user interests, as expressed through the votes they
make, grows, performance of all models improves, but that of \laLDA
improves much faster. \laLDA correctly predicts more than 30\% of the
votes made by the most active users, as compared to 11\% of the
randomly guess.
\begin{center} 
\begin{tabular}{|c||c|c|c|c|c|c|c|c|}
\hline
\scriptsize {Average} &  \multicolumn{4}{|c|}{2009 Data } & \multicolumn{4}{|c|}{2010 Data}    \\
\cline{2-9}
\scriptsize {Precision} & \scriptsize{All users} & \scriptsize {$\ge$7.5\%} & \scriptsize{$\ge$15\%}  & \scriptsize {$\ge$25\%}
& \scriptsize{All users} & \scriptsize {$\ge$7.5\%} & \scriptsize{$\ge$15\%}  & \scriptsize {$\ge$25\%}  \\
\hline\hline
\scriptsize {random} & \scriptsize{0.0192} & \scriptsize {0.0477} & \scriptsize{0.0617} & \scriptsize {0.1092}
& \scriptsize{0.0111} & \scriptsize {0.03619} & \scriptsize{0.0557} & \scriptsize {0.1054} \\
\hline
\scriptsize {LDA} & \scriptsize{0.0209} & \scriptsize {0.0440} & \scriptsize{0.0621}& \scriptsize {0.1107}
& \scriptsize{0.0182} & \scriptsize {0.0415} & \scriptsize{0.0562} & \scriptsize {0.1117}  \\
\hline
\scriptsize {ITM} & \scriptsize{0.0220} & \scriptsize {0.1100} & \scriptsize{0.1526} & \scriptsize {0.2693}
& \scriptsize{0.0244} & \scriptsize {0.1363} & \scriptsize{0.1763}   & \scriptsize {0.2370} \\
\hline
\scriptsize {\laLDA} & \scriptsize{0.0224} & \scriptsize {\textbf{0.1164}} & \scriptsize{\textbf{0.1677}}  & \scriptsize {\textbf{0.3204}}
& \scriptsize{0.0376} & \scriptsize {\textbf{0.1368}} & \scriptsize{\textbf{0.1881}} &  \scriptsize {\textbf{0.3154}}   \\
\hline
\hline
\scriptsize {Submitter} & \scriptsize{0.0379} & \scriptsize {0.0873} & \scriptsize{0.1138}& \scriptsize {0.1517}
& \scriptsize{0.0283} & \scriptsize {0.0483} & \scriptsize{0.0746}  & \scriptsize {0.1257}  \\
\hline
\scriptsize {Max} & \scriptsize{\textbf{0.0789}} & \scriptsize {0.0964} & \scriptsize{0.1240}& \scriptsize {0.1707}
& \scriptsize{\textbf{0.0702}} & \scriptsize {0.0733} & \scriptsize{0.1080}  & \scriptsize {0.1616}  \\
\hline 			
\scriptsize {ITM+Submitter} & \scriptsize{0.0241} & \scriptsize {0.0904} & \scriptsize{0.1311} & \scriptsize {0.1889}
& \scriptsize{0.0381} & \scriptsize {0.0845} & \scriptsize{0.1121}   & \scriptsize {0.1816} \\
\hline
\scriptsize {ITM+Max} & \scriptsize{0.0257} & \scriptsize {0.0977} & \scriptsize{0.1471} & \scriptsize {0.2365}
& \scriptsize{0.0482} & \scriptsize {0.1243} & \scriptsize{0.1645}  & \scriptsize {0.2436} \\
\hline
\end{tabular}
\end{center}

One may ask whether a
simple attention allocation heuristic could predict votes as well
as \laLDA, but at a reduced computational cost.
We answer this question by presenting results of four experiments studying
the effect of the influence heuristic on the prediction task. In the
first experiment, predicted votes for each user are sorted based the
influence of the \emph{submitter}, the first user to post the story on
Digg. In the second experiment, they are sorted based on the influence
of the most influential (\emph{max}) voter. The third experiment
investigates the effect of including either influence heuristic into
the ITM model. In this case, the vote probability given by
Eq.~\ref{eq:predict-itm} is multiplied by relative influence (with
respect to the most influential user in the network) of
the \emph{submitter} or \emph{max} voter.  When there is little
information to learn user interests, using a simple heuristic that a
user votes for a story if a very influential user recommended it, works
well to predict votes, three to four times better than
random guess. However, as \laLDA receives more data about
user interests, it is able to learn a model that outperforms the simpler influence-based models.

\section{Conclusion}
Traditional topic models have been extended to a networked
setting to model hyperlinks between
documents~\cite{citeulike:2914131},
and the varying vocabularies and styles of different authors~\cite{RosenZvtheauthortopic}.
Collaborative filtering methods
examine item recommendations made by many users to discover their
preferences and recommend new items that were liked by similar
users~(\cite{Sarwar01itembasedcollaborative},\cite{Lauw12}) and  improve the explanatory power of recommendations by extending LDA~\cite{WangB11}. 

We introduced \laLDA, a novel hidden topic model 
that takes into account social media users' limited attention.
Our work demonstrates the importance of modeling psychological
factors, such as attention, in social media analysis. These results may apply beyond social media and point to the
fundamental role that psychosocial and cognitive factors play in
social communication. People do not have infinite time and patience to
read all status updates or scientific articles on topics they are
interested in, see all the movies or read all the books. Attention
acts as an ``information bottleneck,'' selecting a small fraction of
available input for further processing. Since human attention is
finite, the mechanisms that guide it become ever more
important. Uncovering the factors that guide attention
will be the focus of our future work.

\bibliographystyle{abbrv}
\bibliography{sigproc}

\end{document}